\documentclass[reprint,nofootinbib,amsmath,amssymb,aps,preprintnumbers,superscriptaddress]{revtex4-1}

\usepackage{bm}
\usepackage{braket}
\usepackage{cases} 
\usepackage{color}
\usepackage{upgreek}
\usepackage{ulem}
\usepackage{cancel}
\usepackage{comment}
\usepackage{empheq}
\usepackage{here}
\usepackage{hyperref}
\hypersetup{colorlinks=true, allcolors=blue}

\begin{document}

\preprint{KOBE-COSMO-26-09}

\title{X-ray signals converted from high-frequency gravitational waves emitted by spinning light primordial black hole dark matter}

\author{Asuka Ito}
\affiliation{Department of Physics, Kobe University, Kobe 657-8501, Japan}

\author{Kazunori Kohri}
\affiliation{National Astronomical Observatory of Japan, 2-21-1 Osawa, Mitaka, Tokyo 181-8588, Japan}
\affiliation{Department of Astronomy, The University of Tokyo, Bunkyo-ku, Hongo, Tokyo 113-0033, Japan}
\affiliation{SOKENDAI, 2-21-1 Osawa, Mitaka, Tokyo 181-8588, Japan}
\affiliation{Theory Center, IPNS, KEK, 1-1 Oho, Tsukuba, Ibaraki 305-0801, Japan}
\affiliation{Kavli IPMU (WPI), UTIAS, The University of Tokyo, Kashiwa, Chiba 277-8583, Japan}

\begin{abstract}
We investigate the detectability of high-frequency gravitational waves from the superradiance of 
light primordial black hole dark matter through photons converted in the Galactic magnetic field.
We find that the signal is significantly enhanced in the X-ray frequency range around $10^{18}$Hz.
For clustered initial conditions, future X-ray observations may detect the converted photons from primordial black holes in the mass range 
$10^{-15}M_{\odot} \sim 10^{-13}M_{\odot}$.
Our results indicate that future X-ray observations could provide a new probe 
of light primordial black hole dark matter through 
high-frequency gravitational waves.
\end{abstract}

\maketitle

\section{Introduction}
Primordial black holes (PBHs) are well-motivated candidates for dark matter (DM)~\cite{Carr:2009jm,Carr:2016drx,Carr:2020gox}. (See also some reviews~\cite{Sasaki:2018dmp,Green:2020jor,Carr:2021bzv,escriva:2022duf,Carr:2025kdk,Carr:2026hot}.) 
Current observational constraints still allow light PBHs with masses in the range $10^{-15}M_{\odot} \sim 10^{-11}M_{\odot}$ to account for the entire DM abundance~\cite{Carr:2020gox,Franciolini:2022htd}. 
It is therefore important to develop observational methods to probe this PBH DM scenario. 
One possible approach is to utilize high-frequency GWs~\cite{Franciolini:2022htd}. 
Indeed, scalar-induced GWs could provide a signal of light PBH DM~\cite{Saito:2008jc,Kohri:2018awv,Cai:2018dig,Inomata:2018epa,Cheong:2022gfc,Furuta:2025fbh}. 
Moreover, since the abundance of light PBHs is large and a certain fraction of them can form binaries, GWs emitted by light PBH binaries can serve as probes through either transient signals or a stochastic background~\cite{Furuta:2025fbh,Franciolini:2022htd,Carr:2009jm,Sasaki:2018dmp,Carr:2020gox}.

Interestingly, spinning PBHs can also emit coherent continuous GWs in the presence of a bosonic field, such as an axion-like particle.
Although PBHs formed during radiation domination are generally expected to have negligible spins at 
formation~\cite{Harada:2020pzb,Chiba:2017rvs}, 
rapidly spinning PBHs can arise through other formation or evolution scenarios. 
For example, PBHs formed during an early matter-dominated era can have large 
spins~\cite{Ye:2025wif,Harada:2017fjm}, 
while PBHs may also acquire spin through baryonic accretion~\cite{DeLuca:2020bjf,DeLuca:2020qqa}. 
Here, we consider PBHs that acquire spin through mergers in the present-day Universe.
Such spinning light PBHs can undergo superradiance in the presence of a bosonic field and emit high-frequency GWs. 
Therefore, high-frequency GWs provide a promising probe of light PBH DM, motivating the development of detection methods~\cite{Aggarwal:2025noe,Ito:2019wcb,Ito:2020wxi,Berlin:2021txa,Ito:2022rxn,Ito:2023bnu,Berlin:2023grv,Bringmann:2023gba,Kanno:2023whr,Matsuo:2025blj,Ito:2025mgm,Cai:2025fpe,Pshirkov:2009sf,Dolgov:2012be,Domcke:2020yzq,Ramazanov:2023nxz,Liu:2023mll,Ito:2023fcr,Ito:2023nkq,Dandoy:2024oqg,He:2023xoh,Lella:2024dus,McDonald:2024nxj,Kushwaha:2025mia,Li:2025eoo,Amaral:2026bef,Baker:2026dqf,Matsuo:2026lvo,Arvanitaki:2012cn}.

In this study, we investigate the detectability of the high-frequency GWs from the superradiance of light PBH DM 
through photons converted in the Galactic magnetic field. 
Recently, a similar idea was discussed for radio signals from PBHs with masses $10^{-5}M_{\odot} \sim 10^{-3}M_{\odot}$, which can constitute a fraction of DM~\cite{Baker:2026dqf}. 
Our study is complementary to that work, focusing on the mass range $10^{-15}M_{\odot}\sim 10^{-11}M_{\odot}$, where PBHs can account for all of DM and the resulting signal lies in the X-ray band.

\section{Superradiant GWs from light PBHs}
A bosonic field such as an axion-like particle 
can cause superradiance of spinning PBHs
when its Compton wavelength is comparable to the size of the PBHs, 
resulting in the emission of coherent GWs whose angular frequency is given by twice the boson mass 
$m_{{\rm b}}$~\cite{Dicke:1954zz,Brito:2015oca}.
Using the parameter $\alpha = 2 G m_{{\rm b}} m_{{\rm PBH}}$, where $G$ is the gravitational constant and 
$m_{{\rm PBH}}$ is the PBH mass, the frequency of the GWs from the superradiance is 
\begin{eqnarray}
  f &\sim& 4.8 \times 10^{17} {\rm Hz}    \left(\frac{m_{{\rm b}}}{{\rm keV}}\right) \\
     &=& 3.2 \times 10^{17} {\rm Hz}    \left( \frac{\alpha}{0.1} \right) 
       \left(\frac{2 m_{{\rm PBH}}}{2 \times 10^{-14} M_{\odot}}\right)^{-1} . 
       \label{fre}
\end{eqnarray}
We see that high-frequency GWs can be a potential probe of light PBH DM~\cite{Franciolini:2022htd}.
We also find that a bosonic field with a keV-scale mass is required in order to have such superradiance.
Interestingly, the possible existence of a spin-0 field with a keV-scale mass has also been independently suggested 
by several observations~\cite{Jaeckel:2014qea,Conlon:2014xsa,Higaki:2014qua,Boyarsky:2014jta,Takahashi:2020bpq}.
In the following, we focus on a spin-0 field.
The GW amplitude from the superradiance is
\begin{equation}
h_0 \simeq   10^{-38} 
            \left( \frac{2 m_{{\rm PBH}}}{2\times 10^{-14} M_{\odot}} \right)
            \left( \frac{\alpha}{0.1} \right)^7
            \left( \frac{{\rm kpc}}{r} \right)
            \left( \frac{\Delta \chi}{0.5} \right) ,  \label{h}
\end{equation}
and the GW emission continues for the coherence time~\cite{Brito:2015oca,Aggarwal:2025noe}
\begin{equation}
  \tau_c \simeq 8.3 \times 10^{-3} {\rm s} 
                 \left(\frac{2 m_{{\rm PBH}}}{2\times 10^{-14} M_{\odot}}\right)
                 \left( \frac{\alpha}{0.1} \right)^{-15}
                 \left( \frac{\Delta \chi}{0.5} \right)^{-1} ,
\end{equation} 
where $r$ is the distance to the PBH and
$\Delta \chi$ represents the difference between the initial and final values of the dimensionless spin parameter of the PBH.
Since we consider spinning PBHs formed through mergers, 
the above parameters are evaluated using approximately twice the mass of the pre-merger PBHs, $m_{{\rm PBH}}$.
The event rate of PBH superradiance is expected to be equivalent to that of PBH mergers.
In order to estimate the event rate, we assume that PBHs make up 100\% 
of the DM abundance, which is allowed in the mass range around $10^{-15}M_{\odot} \sim 10^{-11}M_{\odot}$, and that
their mass distribution is narrow.
In this case, as discussed in~\cite{Franciolini:2022htd,Hutsi:2020sol}, 
the event rate per unit time and unit volume in the 
Milky Way can be estimated as
\begin{equation}
  R_{{\rm PBH}} \simeq 8.8 \times 10^{3}\,  {\rm kpc}^{-3} {\rm yr}^{-1}
               \left( \frac{\bar{\rho}_{{\rm G}} / \rho_{{\rm DM}}}{9.3\times 10^5} \right)
               \left( \frac{m_{{\rm PBH}}}{10^{-14} M_{\odot}} \right)^{-\frac{32}{37}}  .    \label{R}        
\end{equation}
In deriving the event rate, we used the averaged Galactic DM density
$\bar{\rho}_{\rm G}$ (see Appendix~\ref{a} for details), assuming a narrow field of view (FoV) directed toward a region around the Galactic center. 
The quantity $\rho_{\rm DM}=1.3\times10^{-6}\,{\rm GeV}/{\rm cm}^3$ denotes the cosmological mean DM density~\cite{Planck:2018vyg}.
The above event rate is derived assuming a non-clustered initial condition for PBH formation.
In contrast, a clustered initial spatial distribution, which is expected for non-Gaussian primordial fluctuations, can enhance the present-day merger rate up to~\cite{Franciolini:2022htd,DeLuca:2021hde}
\begin{equation}
    R^{{\rm max}}_{{\rm PBH}} \simeq 6.1 \times 10^{9} 
               \,  {\rm kpc}^{-3} {\rm yr}^{-1}
               \left( \frac{\bar{\rho}_{{\rm G}} / \rho_{{\rm DM}}}{9.3\times 10^5} \right)
               \left( \frac{m_{{\rm PBH}}}{10^{-14} M_{\odot}} 
               \right)^{-1}  .     \label{R2}     
\end{equation}
For a given PBH DM mass, the merger rate lies in the range between Eqs.~(\ref{R}) and~(\ref{R2}).
The accumulated GW background from superradiance following the merger events gives
$h^2\Omega_{{\rm GW}}\simeq (1\text{--}2)\times10^{-8}$ for Eq.~(\ref{R}) and 
$h^2\Omega_{{\rm GW}}\simeq 1.0\times10^{-2}$ for Eq.~(\ref{R2}).
The non-clustered case has a slight frequency dependence, whereas the clustered case does not.
In the next section, we evaluate the detectability of superradiant GWs from PBH DM with X-ray telescopes through their conversion into photons in the Galactic magnetic field.

\section{Graviton to photon conversion in the Milky Way}
It has been observed that magnetic fields are present in the Milky Way Galaxy.
The typical magnetic field strength $B_{\rm G}$ lies in the range of 
$1\mu{\rm G} \sim  10\mu{\rm G}$~\cite{Haverkorn:2014jka,Boulanger:2018zrk}.
There are two main components of the Galactic magnetic field: one is the large-scale magnetic field, which has
a large coherence length and a directional dependence following the Galactic spiral structure,
and the other is the small-scale magnetic field, whose coherence length is of order
$1\mathrm{pc} \sim 100\mathrm{pc}$~\cite{Haverkorn:2004fw,Iacobelli:2013fqa,Haverkorn:2008tb}.
Here, we focus on the large-scale component and model it
by taking the magnitude of its transverse component to be $B_{\rm G} \sim 1\mu{\rm G}$ on average over a Galactic length $l_{\rm G} \sim 20{\rm kpc}$, for simplicity.
The conversion probability of gravitons into photons in the above model is given 
by~\cite{Raffelt:1987im,Ito:2023nkq}
\begin{empheq}[left={P(r)\simeq \empheqlbrace}]{alignat=2}
&\frac{r^2 B_{\rm G}^2}{2M_{\rm pl}^2}
&\quad& \text{for } \  r \leq l_{\rm os}, \notag\\
&\frac{l_{\rm os}^2 B_{\rm G}^2}{2M_{\rm pl}^2}
&\quad& \text{for } \   l_{\rm os} \leq r ,  \label{P}
\end{empheq}
where $r$ denotes the propagation length, namely the distance from the superradiant source to the observer, and the oscillation length $l_{\rm os}$ is defined by
\begin{equation}
  l_{{\rm os}} \simeq \frac{4\omega}{\omega_{p}^{2}}  
               \simeq 20 {\rm kpc} 
                 \left( \frac{f}{2.6\times 10^{18}{\rm Hz}} \right).
            \label{os} 
\end{equation}
The plasma frequency due to the presence of electrons is defined by 
$\omega_{p}=\sqrt{\frac{4\pi\alpha_e n_{e}}{m_{e}}}$,
where $\alpha_e$ is the fine structure constant and $m_e$ is the electron mass.
Here, we assume that the electron density in the Milky Way is of order
$n_e \sim 10^{-2}\mathrm{cm}^{-3}$ on average~\cite{Cordes:2002wz}.
Using the above modeling and parameterizations, 
the conversion probability in the Milky Way can be estimated as
$\frac{(20{\rm kpc})^2 (1\mu {\rm G})^2}{2 M_{{\rm pl}}^2}
\simeq 3 \times 10^{-16}$.
We note that, as discussed in~\cite{Lella:2024dus}, this simple modeling yields a conversion probability comparable to 
that obtained using a more sophisticated model of the Galactic 
magnetic field configuration~\cite{2012ApJ...757...14J}.

We now estimate the number of photons converted from GWs emitted by
superradiant events within a given FoV around the Galactic center.
For an observation with a telescope with effective area $A$, 
the number of observed photons from a single event at distance $r$ is calculated as
$\frac{M_{{\rm pl}}^2 \omega h_0^2(r) \tau_c A}{2}
\times P(r)$.
From Eqs.(\ref{h}) and (\ref{P}), the photon number produced by a single event is independent 
of the distance as long as $r \leq l_{\rm os}$.
Then, an event at a larger distance makes an equal contribution, 
resulting in an enhancement when integrated over the source distribution.
Using the event rate in Eq.~(\ref{R}) (or Eq.~(\ref{R2})),
the total number of photons from all events within the narrow FoV $\Omega_{{\rm ob}}$
around the Galactic center during an observation time $t_{\rm ob}$ is given by
\begin{widetext}
\begin{eqnarray}
N_{\rm tot}(\omega) 
&\simeq&
\Omega_{{\rm ob}} t_{\rm ob}R_{\rm PBH}
\int_0^{l_{\rm G}}
\frac{M_{\rm pl}^2\omega h_0(r)^2\tau_c A}{2}
P(r)
r^2\,dr,
\nonumber\\
&=&
\left\{
\begin{array}{ll}
\displaystyle
\frac{\Omega_{{\rm ob}}t_{\rm ob}\tau_c A\omega
h_0^2(l_{\rm os})l_{\rm os}^2B_{\rm G}^2R_{\rm PBH}}{12}
l_{\rm G}^3, \quad
&
{\rm for} \   l_{\rm G}\leq l_{\rm os},
\\[12pt]
\displaystyle
\frac{\Omega_{{\rm ob}} t_{\rm ob}\tau_c A\omega
h_0^2(l_{\rm os})l_{\rm os}^2B_{\rm G}^2R_{\rm PBH}}{12}
\left[
l_{\rm os}^3
+
3l_{\rm os}^2
\left(
l_{\rm G}-l_{\rm os}
\right)
\right], \quad
&
{\rm for} \  l_{\rm os}<l_{\rm G}.
\end{array}
\right.
\label{Ntot}
\end{eqnarray}
\end{widetext}
%
In the regime $r<l_{{\rm os}}$, more distant events collectively make larger contributions to the total photon number. 
Consequently, the integration is significantly enhanced, particularly when $l_{\rm G}\lesssim l_{{\rm os}}$. 
Eqs.~(\ref{fre}) and~(\ref{os}) show that the condition $l_{\rm G}\lesssim l_{{\rm os}}$ corresponds to frequencies above $\sim 10^{18}{\rm Hz}$, or to PBH masses below $\sim 10^{-14}M_{\odot}$. 
Therefore, superradiant GWs from PBH DM may be observable in the X-ray regime.

\begin{figure}[ht]
\centering
\includegraphics[width=0.50\textwidth]{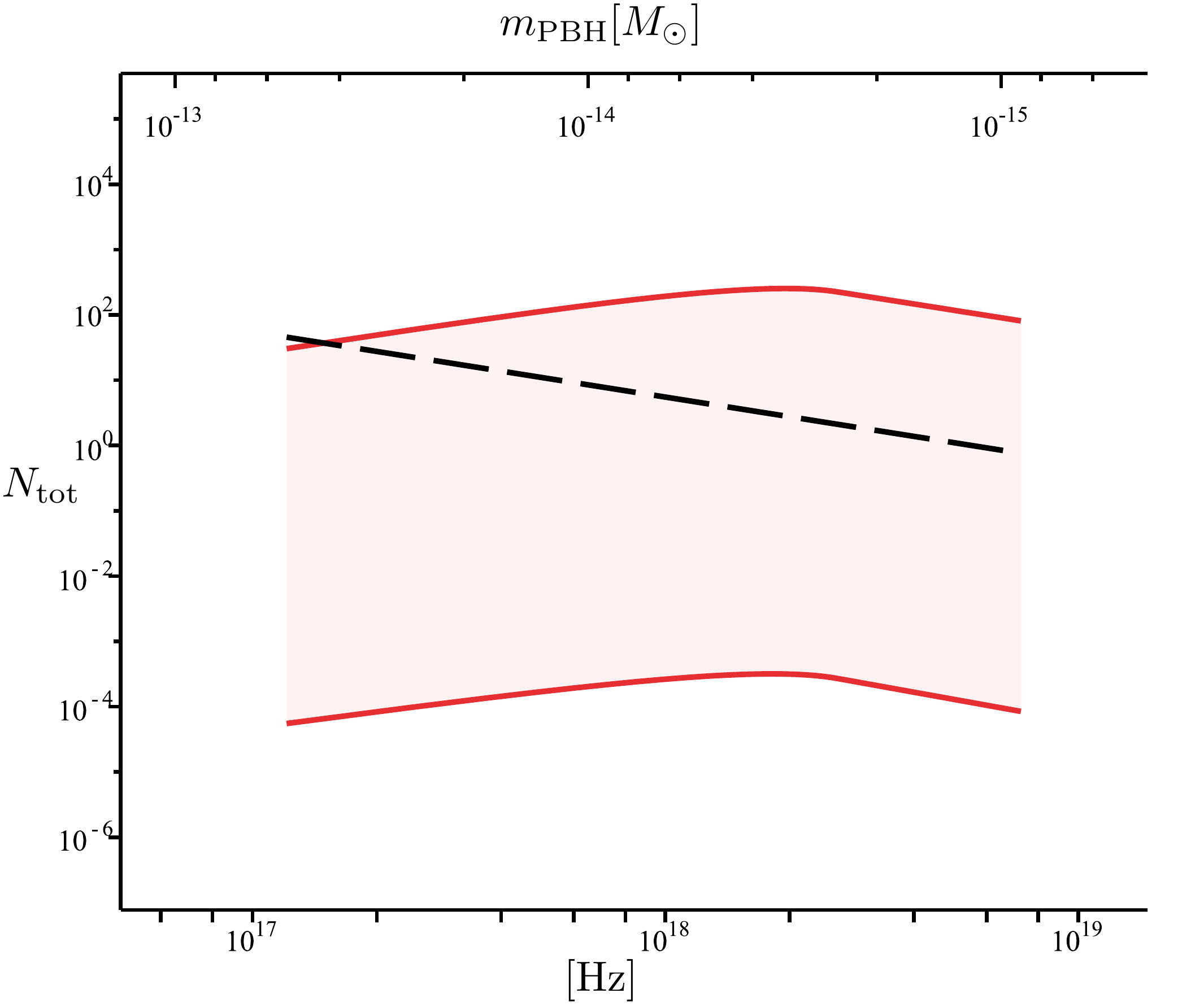}
\caption{The expected photon numbers converted from superradiant GWs emitted by PBH DM are compared with the 
estimated sensitivity for a benchmark X-ray observational setup.
The red shaded region shows the photon numbers during an observation time $t_{\rm ob}=1\,{\rm yr}$, 
assuming an effective area $A=30\,{\rm m}^2$ and a FoV of $30\,{\rm deg}^2$.
The lower and upper edges of the shaded region correspond to the merger rates for the non-clustered and 
clustered initial conditions, Eqs.(\ref{R}) and (\ref{R2}), respectively.
The black dashed line shows the estimated sensitivity of the telescope.} 
\label{zu}
\end{figure}

In Fig.~\ref{zu}, we compare the expected photon number from superradiance of PBH DM with the estimated 
sensitivity for a benchmark X-ray observational setup.
The red shaded region represents the expected photon number converted
from superradiant GWs of PBH DM during an observation time $t_{\rm ob}=1\,{\rm yr}$, 
assuming an effective area $A=30\,{\rm m}^2$ and a FoV of $30\,{\rm deg}^2$.
The lower boundary of the shaded region corresponds to the event rate
for the non-clustered initial condition in Eq.~(\ref{R}), while the
upper boundary corresponds to that for the clustered initial condition
in Eq.~(\ref{R2}).
The black dashed line shows the estimated sensitivity of this X-ray telescope.
We estimate it using the projected Lynx sensitivity,
$1.6\times10^{-19}\,{\rm erg}/{\rm cm}^2/{\rm s}$~\cite{Vikhlinin:2022sde,LynxConceptStudy},
as a reference and scaling it to the observational setup assumed here.
For the energy resolution, we take the optimal value 
$\Delta E/E=4\times10^{-4}$, 
determined by the Doppler broadening due to the Galactic DM velocity dispersion~\cite{Speckhard:2015eva}.
As seen in Fig.~\ref{zu}, the expected photon number increases with frequency up to the scale given by Eq.(\ref{os}), 
and can exceed the estimated sensitivity under the clustered initial condition for PBH masses around 
$10^{-15}M_{\odot} \sim 10^{-13}M_{\odot}$.
This indicates that future X-ray observations may enable us to probe a part of the PBH parameter space.

\section{Conclusion}
We investigated the detectability of high-frequency GWs from the superradiance of light PBH DM by considering photons converted in the Galactic magnetic field.
We found that the signal is significantly enhanced in the X-ray frequency range, 
with a peak around $10^{18}\,{\rm Hz}$, where distant events give an important contribution to the converted photons.
We showed that such photons may be detectable for PBHs in the mass range 
$10^{-15}M_{\odot} \sim 10^{-13}M_{\odot}$ if they constitute all of DM and have clustered initial conditions, 
as expected in scenarios with non-Gaussian primordial fluctuations, with an optimistic setup 
for future X-ray observations.

From the viewpoint of the string axiverse~\cite{Arvanitaki:2009fg,Svrcek:2006yi,Cicoli:2012sz}, 
it may be natural to have spin-0 fields with masses around $100\,{\rm eV} \sim 10\,{\rm keV}$, 
which cause the desired superradiance.
Interestingly, several observations have also suggested the possible existence of an unknown spin-0 field 
in this mass range~\cite{Jaeckel:2014qea,Conlon:2014xsa,Higaki:2014qua,Boyarsky:2014jta,Takahashi:2020bpq}.
Moreover, although we have focused on a spin-0 field, a similar discussion can be applied to bosonic fields with arbitrary spins~\cite{Brito:2015oca,Aggarwal:2025noe}.
Therefore, future X-ray observations could open a new observational window on light PBH DM through superradiant 
high-frequency GWs.

\begin{acknowledgments}
This work was supported by JSPS KAKENHI Grant Numbers JP26K17149 (AI) and JP24K07027 (KK).
\end{acknowledgments}

\appendix

\section{Galactic dark matter density}
\label{a}

Here, we describe the estimate of the averaged Galactic DM density
$\bar{\rho}_{\rm G}$ used in the event rate.
We assume that PBHs follow the Galactic DM distribution and adopt the
Navarro-Frenk-White (NFW) profile,
\begin{equation}
\rho_{\rm NFW}(d)
=
\rho_\odot
\frac{R_\odot}{d}
\left(
\frac{1+R_\odot/d_s}{1+d/d_s}
\right)^2,
\label{eq:NFW}
\end{equation}
where we take~\cite{deSalas:2020hbh}
\begin{equation}
\rho_\odot=0.45\,{\rm GeV}/{\rm cm}^3,
\ \ 
R_\odot=8.2\,{\rm kpc},
\ \ 
d_s=20\,{\rm kpc}.
\end{equation}
Here, $R_\odot$ is the distance between the Solar system and the
Galactic center, and $\rho_\odot$ is the local DM density there.

We consider a narrow FoV directed toward a region around
the Galactic center.
We denote the angular separation between the center of the field of
view and the Galactic center by $\theta_{\rm off}$.
Since the FoV is assumed to be sufficiently narrow, we
neglect the angular variation of the DM density inside the field of
view.
The distance from the Galactic center to a point located at a
line-of-sight distance $r$ from the Solar system is given by
\begin{equation}
d(r,\theta_{\rm off})
=
\sqrt{
R_\odot^2+r^2
-2R_\odot r\cos\theta_{\rm off}
}.
\label{eq:rLOS}
\end{equation}
We define the averaged Galactic DM density within the narrow FoV as
\begin{equation}
\bar{\rho}_{\rm G}
=
\frac{1}{l_{\rm G}^3/3}
\int_0^{l_{\rm G}}
\rho_{\rm NFW}
\left(
d(r,\theta_{\rm off})
\right) r^2 \, dr.
\label{eq:rhoGave}
\end{equation}
The integral in Eq.~(\ref{eq:rhoGave}) can be evaluated analytically as
\begin{widetext}
\begin{eqnarray}
&&\int_0^{l_{\rm G}}
\rho_{\rm NFW}
\left(
d(r,\theta_{\rm off})
\right)  r^2 \, dr  \nonumber \\
&& \hspace{0.2cm}=
\rho_{\odot}R_{\odot}(R_{\odot}+d_s)^2
\Bigg[
\frac{d_s(R_{\odot}^2-d_s^2)}{\Delta^3}
\ln\left(
\frac{
\left(\sqrt{(l_G-c)^2+b^2}+b+k(l_G-c)\right)
\left(R_{\odot}+b+kc\right)
}{
\left(\sqrt{(l_G-c)^2+b^2}+b-k(l_G-c)\right)
\left(R_{\odot}+b-kc\right)
}
\right)
\nonumber\\
&& \hspace{3.3cm}
+\ln\left(
\frac{
l_G-c+\sqrt{(l_G-c)^2+b^2}
}{
R_{\odot}-c
}
\right)
-\frac{
l_G+c+\dfrac{c^2(l_G-c)}{\Delta^2}
}{
d_s+\sqrt{(l_G-c)^2+b^2}
}
+\frac{
c-\dfrac{c^3}{\Delta^2}
}{
d_s+R_{\odot}
}
\Bigg],
\label{eq:NFWLOS}
\end{eqnarray}
\end{widetext}
where $b=R_{\odot} \sin\theta_{{\rm off}}$, $c=R_{\odot} \cos\theta_{{\rm off}}$, $\Delta = \sqrt{d_s^2 - b^2}$,
and $k=\sqrt{(d_s - b)/(d_s + b)}$.

In our analysis, we take
\begin{equation}
\theta_{\rm off}=5^\circ ,
\end{equation}
as a benchmark value.
We note that, when the angular offset is taken in Galactic latitude,
i.e., away from the Galactic plane, it is possible to target a
dark region with a low astrophysical X-ray background.
For $R_\odot=8.2\,{\rm kpc}$, $d_s=20\,{\rm kpc}$ and
$l_{\rm G}=20\,{\rm kpc}$, we obtain
\begin{equation}
\bar{\rho}_{\rm G}
\simeq
1.2\,{\rm GeV}/{\rm cm}^3.
\label{eq:rhoGnum}
\end{equation}
Thus, the Galactic DM distribution gives the enhancement factor
$\bar{\rho}_{\rm G}/\rho_{\rm DM}\simeq 9.3\times10^5$ appearing in
Eqs.~(5) and (6), where
$\rho_{\rm DM}=1.3\times10^{-6}\,{\rm GeV}/{\rm cm}^3$ is the
cosmological mean DM density~\cite{Planck:2018vyg}.

\bibliography{ref}

@article{Carr:2025kdk,
    author = "Carr, Bernard and Kuhnel, Florian",
    title = "{Primordial Black Holes}",
    eprint = "2502.15279",
    archivePrefix = "arXiv",
    primaryClass = "astro-ph.CO",
    month = "2",
    year = "2025"
}

@article{Carr:2021bzv,
    author = "Carr, Bernard and Kuhnel, Florian",
    title = "{Primordial black holes as dark matter candidates}",
    eprint = "2110.02821",
    archivePrefix = "arXiv",
    primaryClass = "astro-ph.CO",
    doi = "10.21468/SciPostPhysLectNotes.48",
    journal = "SciPost Phys. Lect. Notes",
    volume = "48",
    pages = "1",
    year = "2022"
}

@article{Carr:2020gox,
    author = "Carr, Bernard and Kohri, Kazunori and Sendouda, Yuuiti and Yokoyama, Jun'ichi",
    title = "{Constraints on primordial black holes}",
    eprint = "2002.12778",
    archivePrefix = "arXiv",
    primaryClass = "astro-ph.CO",
    reportNumber = "RESCEU-03/20; KEK-Cosmo-249; KEK-TH-2199; IPMU20-0024",
    doi = "10.1088/1361-6633/ac1e31",
    journal = "Rept. Prog. Phys.",
    volume = "84",
    number = "11",
    pages = "116902",
    year = "2021"
}

@article{Franciolini:2022htd,
    author = "Franciolini, Gabriele and Maharana, Anshuman and Muia, Francesco",
    title = "{Hunt for light primordial black hole dark matter with ultrahigh-frequency gravitational waves}",
    eprint = "2205.02153",
    archivePrefix = "arXiv",
    primaryClass = "astro-ph.CO",
    doi = "10.1103/PhysRevD.106.103520",
    journal = "Phys. Rev. D",
    volume = "106",
    number = "10",
    pages = "103520",
    year = "2022"
}

@article{Carr:2009jm,
    author = "Carr, B. J. and Kohri, Kazunori and Sendouda, Yuuiti and Yokoyama, Jun'ichi",
    title = "{New cosmological constraints on primordial black holes}",
    eprint = "0912.5297",
    archivePrefix = "arXiv",
    primaryClass = "astro-ph.CO",
    reportNumber = "RESCEU-31-09, TU-852, YITP-09-112",
    doi = "10.1103/PhysRevD.81.104019",
    journal = "Phys. Rev. D",
    volume = "81",
    pages = "104019",
    year = "2010"
}

@article{Sasaki:2018dmp,
    author = "Sasaki, Misao and Suyama, Teruaki and Tanaka, Takahiro and Yokoyama, Shuichiro",
    title = "{Primordial black holes{\textemdash}perspectives in gravitational wave astronomy}",
    eprint = "1801.05235",
    archivePrefix = "arXiv",
    primaryClass = "astro-ph.CO",
    doi = "10.1088/1361-6382/aaa7b4",
    journal = "Class. Quant. Grav.",
    volume = "35",
    number = "6",
    pages = "063001",
    year = "2018"
}

@article{Brito:2015oca,
    author = "Brito, Richard and Cardoso, Vitor and Pani, Paolo",
    title = "{Superradiance}: {New Frontiers in Black Hole
Physics}",
    eprint = "1501.06570",
    archivePrefix = "arXiv",
    primaryClass = "gr-qc",
    doi = "10.1007/978-3-319-19000-6",
    journal = "Lect. Notes Phys.",
    volume = "906",
    pages = "pp.1--237",
    year = "2015"
}

@article{Hutsi:2020sol,
    author = {H{\"u}tsi, Gert and Raidal, Martti and Vaskonen, Ville and Veerm{\"a}e, Hardi},
    title = "{Two populations of LIGO-Virgo black holes}",
    eprint = "2012.02786",
    archivePrefix = "arXiv",
    primaryClass = "astro-ph.CO",
    doi = "10.1088/1475-7516/2021/03/068",
    journal = "JCAP",
    volume = "03",
    pages = "068",
    year = "2021"
}

@article{DeLuca:2021hde,
    author = "De Luca, Valerio and Franciolini, Gabriele and Pani, Paolo and Riotto, Antonio",
    title = "{The minimum testable abundance of primordial black holes at future gravitational-wave detectors}",
    eprint = "2106.13769",
    archivePrefix = "arXiv",
    primaryClass = "astro-ph.CO",
    doi = "10.1088/1475-7516/2021/11/039",
    journal = "JCAP",
    volume = "11",
    pages = "039",
    year = "2021"
}

@ARTICLE{2012ApJ...757...14J,
       author = {{Jansson}, Ronnie and {Farrar}, Glennys R.},
        title = "{A New Model of the Galactic Magnetic Field}",
      journal = {\apj},
         year = 2012,
        month = sep,
       volume = {757},
       number = {1},
          eid = {14},
        pages = {14},
          doi = {10.1088/0004-637X/757/1/14},
archivePrefix = {arXiv},
       eprint = {1204.3662},
 primaryClass = {astro-ph.GA},
       adsurl = {https://ui.adsabs.harvard.edu/abs/2012ApJ...757...14J}
}

@article{Arvanitaki:2009fg,
    author = "Arvanitaki, Asimina and Dimopoulos, Savas and Dubovsky, Sergei and Kaloper, Nemanja and March-Russell, John",
    title = "{String Axiverse}",
    eprint = "0905.4720",
    archivePrefix = "arXiv",
    primaryClass = "hep-th",
    doi = "10.1103/PhysRevD.81.123530",
    journal = "Phys. Rev. D",
    volume = "81",
    pages = "123530",
    year = "2010"
}

@article{Svrcek:2006yi,
    author = "Svrcek, Peter and Witten, Edward",
    title = "{Axions In String Theory}",
    eprint = "hep-th/0605206",
    archivePrefix = "arXiv",
    reportNumber = "SLAC-PUB-11894",
    doi = "10.1088/1126-6708/2006/06/051",
    journal = "JHEP",
    volume = "06",
    pages = "051",
    year = "2006"
}

@article{Cicoli:2012sz,
    author = "Cicoli, Michele and Goodsell, Mark and Ringwald, Andreas",
    title = "{The type IIB string axiverse and its low-energy phenomenology}",
    eprint = "1206.0819",
    archivePrefix = "arXiv",
    primaryClass = "hep-th",
    reportNumber = "DESY-12-058, CERN-PH-TH-2012-153",
    doi = "10.1007/JHEP10(2012)146",
    journal = "JHEP",
    volume = "10",
    pages = "146",
    year = "2012"
}

@article{Jaeckel:2014qea,
    author = "Jaeckel, Joerg and Redondo, Javier and Ringwald, Andreas",
    title = "{3.55 keV hint for decaying axionlike particle dark matter}",
    eprint = "1402.7335",
    archivePrefix = "arXiv",
    primaryClass = "hep-ph",
    reportNumber = "DESY-14-023, MPP-2014-41",
    doi = "10.1103/PhysRevD.89.103511",
    journal = "Phys. Rev. D",
    volume = "89",
    pages = "103511",
    year = "2014"
}

@article{Conlon:2014xsa,
    author = "Conlon, Joseph P. and Day, Francesca V.",
    title = "{3.55 keV photon lines from axion to photon conversion in the Milky Way and M31}",
    eprint = "1404.7741",
    archivePrefix = "arXiv",
    primaryClass = "hep-ph",
    doi = "10.1088/1475-7516/2014/11/033",
    journal = "JCAP",
    volume = "11",
    pages = "033",
    year = "2014"
}

@article{Higaki:2014qua,
    author = "Higaki, Tetsutaro and Kitajima, Naoya and Takahashi, Fuminobu",
    title = "{Hidden axion dark matter decaying through mixing with QCD axion and the 3.5 keV X-ray line}",
    eprint = "1408.3936",
    archivePrefix = "arXiv",
    primaryClass = "hep-ph",
    reportNumber = "TU-979, IPMU14-0269",
    doi = "10.1088/1475-7516/2014/12/004",
    journal = "JCAP",
    volume = "12",
    pages = "004",
    year = "2014"
}

@article{Boyarsky:2014jta,
    author = "Boyarsky, Alexey and Ruchayskiy, Oleg and Iakubovskyi, Dmytro and Franse, Jeroen",
    title = "{Unidentified Line in X-Ray Spectra of the Andromeda Galaxy and Perseus Galaxy Cluster}",
    eprint = "1402.4119",
    archivePrefix = "arXiv",
    primaryClass = "astro-ph.CO",
    doi = "10.1103/PhysRevLett.113.251301",
    journal = "Phys. Rev. Lett.",
    volume = "113",
    pages = "251301",
    year = "2014"
}

@article{Takahashi:2020bpq,
    author = "Takahashi, Fuminobu and Yamada, Masaki and Yin, Wen",
    title = "{XENON1T Excess from Anomaly-Free Axionlike Dark Matter and Its Implications for Stellar Cooling Anomaly}",
    eprint = "2006.10035",
    archivePrefix = "arXiv",
    primaryClass = "hep-ph",
    reportNumber = "TU-1104, IPMU20-0069",
    doi = "10.1103/PhysRevLett.125.161801",
    journal = "Phys. Rev. Lett.",
    volume = "125",
    number = "16",
    pages = "161801",
    year = "2020"
}

@article{Vikhlinin:2022sde,
    author = {Vikhlinin, Alexey and {\"O}zel, Feryal and Gaskin, Jessica},
    collaboration = "LYNX Team",
    title = "{LYNX X-ray Observatory Concept Study Report}",
    year = "2022"
}

@misc{LynxConceptStudy,
  title        = "{Lynx Concept Study}",
  institution  = "{NASA Marshall Space Flight Center}",
  url          = "https://wwwastro.msfc.nasa.gov/lynx/",
  note         = "Accessed: 2026-07-09"
}

@article{Aggarwal:2025noe,
    author = "Aggarwal, Nancy and others",
    title = "{Challenges and opportunities of gravitational-wave searches above 10 kHz}",
    eprint = "2501.11723",
    archivePrefix = "arXiv",
    primaryClass = "gr-qc",
    reportNumber = "CERN-TH-2025-014, DESY-25-007",
    doi = "10.1007/s41114-025-00060-5",
    journal = "Living Rev. Rel.",
    volume = "28",
    number = "1",
    pages = "10",
    year = "2025"
}

@article{Ito:2019wcb,
    author = "Ito, Asuka and Ikeda, Tomonori and Miuchi, Kentaro and Soda, Jiro",
    title = "{Probing GHz gravitational waves with graviton--magnon resonance}",
    eprint = "1903.04843",
    archivePrefix = "arXiv",
    primaryClass = "gr-qc",
    doi = "10.1140/epjc/s10052-020-7735-y",
    journal = "Eur. Phys. J. C",
    volume = "80",
    number = "3",
    pages = "179",
    year = "2020"
}

@article{Ito:2020wxi,
    author = "Ito, Asuka and Soda, Jiro",
    title = "{A formalism for magnon gravitational wave detectors}",
    eprint = "2004.04646",
    archivePrefix = "arXiv",
    primaryClass = "gr-qc",
    doi = "10.1140/epjc/s10052-020-8092-6",
    journal = "Eur. Phys. J. C",
    volume = "80",
    number = "6",
    pages = "545",
    year = "2020"
}

@article{Berlin:2021txa,
    author = "Berlin, Asher and Blas, Diego and D'Agnolo, Raffaele Tito and Ellis, Sebastian A. R. and Harnik, Roni and Kahn, Yonatan and Schütte-Engel, Jan",
    title = "{Detecting High-Frequency Gravitational Waves with Microwave Cavities}",
    eprint = "2112.11465",
    archivePrefix = "arXiv",
    primaryClass = "hep-ph",
    doi = "10.1103/PhysRevD.105.116011",
    journal = "Phys. Rev. D",
    volume = "105",
    number = "11",
    pages = "116011",
    year = "2022"
}

@article{Ito:2022rxn,
    author = "Ito, Asuka and Soda, Jiro",
    title = "{Exploring High Frequency Gravitational Waves with Magnons}",
    eprint = "2212.04094",
    archivePrefix = "arXiv",
    primaryClass = "gr-qc",
    doi = "10.1140/epjc/s10052-023-11876-2",
    journal = "Eur. Phys. J. C",
    volume = "83",
    number = "8",
    pages = "766",
    year = "2023"
}

@article{Ito:2023bnu,
    author = "Ito, Asuka and Kitano, Ryuichiro",
    title = "{Macroscopic Quantum Response to Gravitational Waves}",
    eprint = "2309.02992",
    archivePrefix = "arXiv",
    primaryClass = "gr-qc",
    doi = "10.1088/1475-7516/2024/04/068",
    journal = "JCAP",
    volume = "04",
    pages = "068",
    year = "2024"
}

@article{Berlin:2023grv,
    author = "Berlin, Asher and Blas, Diego and D'Agnolo, Raffaele Tito and Ellis, Sebastian A. R. and Harnik, Roni and Kahn, Yonatan and Schütte-Engel, Jan and Wentzel, Michael",
    title = "{Electromagnetic cavities as mechanical bars for gravitational waves}",
    eprint = "2303.01518",
    archivePrefix = "arXiv",
    primaryClass = "hep-ph",
    doi = "10.1103/PhysRevD.108.084058",
    journal = "Phys. Rev. D",
    volume = "108",
    number = "8",
    pages = "084058",
    year = "2023"
}

@article{Bringmann:2023gba,
    author = "Bringmann, Torsten and Domcke, Valerie and Fuchs, Elina and Kopp, Joachim",
    title = "{High-frequency gravitational wave detection via optical frequency modulation}",
    eprint = "2304.10579",
    archivePrefix = "arXiv",
    primaryClass = "hep-ph",
    doi = "10.1103/PhysRevD.108.L061303",
    journal = "Phys. Rev. D",
    volume = "108",
    number = "6",
    pages = "L061303",
    year = "2023"
}

@article{Kanno:2023whr,
    author = "Kanno, Sugumi and Soda, Jiro and Taniguchi, Akira",
    title = "{Search for high-frequency gravitational waves with Rydberg atoms}",
    eprint = "2311.03890",
    archivePrefix = "arXiv",
    primaryClass = "gr-qc",
    doi = "10.1140/epjc/s10052-024-13736-z",
    journal = "Eur. Phys. J. C",
    volume = "85",
    number = "1",
    pages = "31",
    year = "2025"
}

@article{Matsuo:2025blj,
    author = "Matsuo, Himeka and Ito, Asuka",
    title = "{Graviton-photon conversion in blazar jets as a probe of high-frequency gravitational waves}",
    eprint = "2505.08457",
    archivePrefix = "arXiv",
    primaryClass = "gr-qc",
    reportNumber = "KOBE-COSMO-25-07",
    doi = "10.1088/1475-7516/2025/10/061",
    journal = "JCAP",
    volume = "10",
    pages = "061",
    year = "2025"
}

@article{Ito:2025mgm,
    author = "Ito, Asuka and Kitano, Ryuichiro and Nakano, Wakutaka and Takai, Ryoto",
    title = "{Quantum sensing of high-frequency gravitational waves with ion crystals}",
    eprint = "2512.19053",
    archivePrefix = "arXiv",
    primaryClass = "gr-qc",
    doi = "10.1088/1475-7516/2026/05/039",
    journal = "JCAP",
    volume = "05",
    pages = "039",
    year = "2026"
}

@article{Cai:2025fpe,
    author = "Cai, Yi-Fu and Visinelli, Luca and Yan, Sheng-Feng",
    title = "{Atomic Quantum Sensors for High-Frequency Gravitational Wave Searches}",
    eprint = "2510.15031",
    archivePrefix = "arXiv",
    primaryClass = "hep-ph",
    doi = "10.1016/j.dark.2026.102374",
    journal = "Phys. Dark Univ.",
    volume = "53",
    pages = "102374",
    year = "2026"
}

@article{Pshirkov:2009sf,
    author = "Pshirkov, M. S. and Baskaran, D.",
    title = "{Limits on high-frequency gravitational wave background from its interplay with large scale magnetic fields}",
    eprint = "0903.4160",
    archivePrefix = "arXiv",
    primaryClass = "gr-qc",
    doi = "10.1103/PhysRevD.80.042002",
    journal = "Phys. Rev. D",
    volume = "80",
    pages = "042002",
    year = "2009"
}

@article{Dolgov:2012be,
    author = "Dolgov, Alexander D. and Ejlli, Damian",
    title = "{Conversion of relic gravitational waves into photons in cosmological magnetic fields}",
    eprint = "1211.0500",
    archivePrefix = "arXiv",
    primaryClass = "gr-qc",
    doi = "10.1088/1475-7516/2012/12/003",
    journal = "JCAP",
    volume = "12",
    pages = "003",
    year = "2012"
}

@article{Domcke:2020yzq,
    author = "Domcke, Valerie and Garcia-Cely, Camilo",
    title = "{Potential of Radio Telescopes as High-Frequency Gravitational Wave Detectors}",
    eprint = "2006.01161",
    archivePrefix = "arXiv",
    primaryClass = "astro-ph.CO",
    doi = "10.1103/PhysRevLett.126.021104",
    journal = "Phys. Rev. Lett.",
    volume = "126",
    number = "2",
    pages = "021104",
    year = "2021"
}

@article{Ramazanov:2023nxz,
    author = "Ramazanov, Sabir and Samanta, Rome and Trenkler, Georg and Urban, Federico R.",
    title = "{Shimmering gravitons in the gamma-ray sky}",
    eprint = "2304.11222",
    archivePrefix = "arXiv",
    primaryClass = "astro-ph.CO",
    doi = "10.1088/1475-7516/2023/06/019",
    journal = "JCAP",
    volume = "06",
    pages = "019",
    year = "2023"
}

@article{Liu:2023mll,
    author = "Liu, Tao and Ren, Jing and Zhang, Chen",
    title = "{Limits on High-Frequency Gravitational Waves in Planetary Magnetospheres}",
    eprint = "2305.01832",
    archivePrefix = "arXiv",
    primaryClass = "hep-ph",
    doi = "10.1103/PhysRevLett.132.131402",
    journal = "Phys. Rev. Lett.",
    volume = "132",
    number = "13",
    pages = "131402",
    year = "2024"
}

@article{Ito:2023fcr,
    author = "Ito, Asuka and Kohri, Kazunori and Nakayama, Kazunori",
    title = "{Probing high frequency gravitational waves with pulsars}",
    eprint = "2305.13984",
    archivePrefix = "arXiv",
    primaryClass = "gr-qc",
    doi = "10.1103/PhysRevD.109.063026",
    journal = "Phys. Rev. D",
    volume = "109",
    number = "6",
    pages = "063026",
    year = "2024"
}

@article{Ito:2023nkq,
    author = "Ito, Asuka and Kohri, Kazunori and Nakayama, Kazunori",
    title = "{Gravitational Wave Search through Electromagnetic Telescopes}",
    eprint = "2309.14765",
    archivePrefix = "arXiv",
    primaryClass = "gr-qc",
    doi = "10.1093/ptep/ptae004",
    journal = "PTEP",
    volume = "2024",
    number = "2",
    pages = "023E03",
    year = "2024"
}

@article{Dandoy:2024oqg,
    author = "Dandoy, Virgile and Bertólez-Martínez, Toni and Costa, Francesco",
    title = "{High Frequency Gravitational Wave Bounds from Galactic Neutron Stars}",
    eprint = "2402.14092",
    archivePrefix = "arXiv",
    primaryClass = "gr-qc",
    doi = "10.1088/1475-7516/2024/12/023",
    journal = "JCAP",
    volume = "12",
    pages = "023",
    year = "2024"
}

@article{He:2023xoh,
    author = "He, Yuanshuai and Giri, Sambit K. and Sharma, Ramkishor and Mtchedlidze, Shota and Georgiev, Ivan",
    title = "{Inverse Gertsenshtein effect as a probe of high-frequency gravitational waves}",
    eprint = "2312.17636",
    archivePrefix = "arXiv",
    primaryClass = "astro-ph.CO",
    doi = "10.1088/1475-7516/2024/05/051",
    journal = "JCAP",
    volume = "05",
    pages = "051",
    year = "2024"
}

@article{Lella:2024dus,
    author = "Lella, Alessandro and Calore, Francesca and Carenza, Pierluca and Mirizzi, Alessandro",
    title = "{Constraining gravitational-wave backgrounds from conversions into photons in the Galactic magnetic field}",
    eprint = "2406.17853",
    archivePrefix = "arXiv",
    primaryClass = "hep-ph",
    doi = "10.1103/PhysRevD.110.083042",
    journal = "Phys. Rev. D",
    volume = "110",
    number = "8",
    pages = "083042",
    year = "2024"
}

@article{McDonald:2024nxj,
    author = "McDonald, Jamie I. and Ellis, Sebastian A. R.",
    title = "{Resonant Conversion of Gravitational Waves in Neutron Star Magnetospheres}",
    eprint = "2406.18634",
    archivePrefix = "arXiv",
    primaryClass = "hep-ph",
    doi = "10.1103/PhysRevD.110.103003",
    journal = "Phys. Rev. D",
    volume = "110",
    number = "10",
    pages = "103003",
    year = "2024"
}

@article{Kushwaha:2025mia,
    author = "Kushwaha, Ashu and Jain, Rajeev Kumar",
    title = "{Constraining circular polarization of high-frequency gravitational waves with CMB}",
    eprint = "2502.12517",
    archivePrefix = "arXiv",
    primaryClass = "astro-ph.CO",
    doi = "10.1103/PhysRevD.112.L021301",
    journal = "Phys. Rev. D",
    volume = "112",
    number = "2",
    pages = "L021301",
    year = "2025"
}

@article{Li:2025eoo,
    author = "Li, Jian-Kang and Hong, Wei and Zhang, Tong-Jie",
    title = "{Polarization Properties of the Electromagnetic Response to High-frequency Gravitational Wave}",
    eprint = "2504.13115",
    archivePrefix = "arXiv",
    primaryClass = "astro-ph.CO",
    journal = "Astrophys. J.",
    volume = "985",
    pages = "137",
    year = "2025"
}

@article{Amaral:2026bef,
    author = "Amaral, Dorian and others",
    title = "{Global detector network to search for high-frequency gravitational waves (GravNet): conceptual design}",
    eprint = "2603.24645",
    archivePrefix = "arXiv",
    primaryClass = "astro-ph.IM",
    month = "3",
    year = "2026"
}

@article{Baker:2026dqf,
    author = "Baker, Ethan and Liu, Hongwan",
    title = "{Radio Emission from High-Frequency Gravitational Wave Point Sources}",
    eprint = "2606.09968",
    archivePrefix = "arXiv",
    primaryClass = "hep-ph",
    year = "2026"
}

@article{Matsuo:2026lvo,
    author = "Matsuo, Himeka and Ito, Asuka and Kohri, Kazunori and Suyama, Teruaki and Tomomatsu, Ryutaro",
    title = "{Graviton Floor}",
    eprint = "2606.19757",
    archivePrefix = "arXiv",
    primaryClass = "gr-qc",
    year = "2026"
}

@article{Arvanitaki:2012cn,
    author = "Arvanitaki, Asimina and Geraci, Andrew A.",
    title = "{Detecting high-frequency gravitational waves with optically-levitated sensors}",
    eprint = "1207.5320",
    archivePrefix = "arXiv",
    primaryClass = "gr-qc",
    doi = "10.1103/PhysRevLett.110.071105",
    journal = "Phys. Rev. Lett.",
    volume = "110",
    number = "7",
    pages = "071105",
    year = "2013"
}

@article{Haverkorn:2014jka,
    author = "Haverkorn, Marijke",
    title = "{Magnetic Fields in the Milky Way}",
    eprint = "1406.0283",
    archivePrefix = "arXiv",
    primaryClass = "astro-ph.GA",
    month = "6",
    year = "2014"
}

@article{Boulanger:2018zrk,
    author = "Boulanger, Fran{\c{c}}ois and others",
    title = "{IMAGINE: A comprehensive view of the interstellar medium, Galactic magnetic fields and cosmic rays}",
    eprint = "1805.02496",
    archivePrefix = "arXiv",
    primaryClass = "astro-ph.GA",
    doi = "10.1088/1475-7516/2018/08/049",
    journal = "JCAP",
    volume = "08",
    pages = "049",
    year = "2018"
}

@article{Haverkorn:2004fw,
    author = "Haverkorn, Marijke and Gaensler, B. M. and McClure-Griffiths, N. M. and Dickey, John M. and Green, Anne J.",
    title = "{Magnetic fields and ionized gas in the inner galaxy: An Outer scale for turbulence and the possible role of HII regions}",
    eprint = "astro-ph/0403655",
    archivePrefix = "arXiv",
    doi = "10.1086/421341",
    journal = "Astrophys. J.",
    volume = "609",
    pages = "776--784",
    year = "2004"
}

\end{document}